\documentclass[11pt,a4paper]{article}
\usepackage{amsmath,amssymb,mathrsfs, bm, graphicx}

  \newcommand{\sla}{\hspace{-0.5em} /}

\begin{document}

\begin{center}
\Large
{\bf Fluctuations of the chiral condensate and quasi-particle spectra
near phase transition}

\vspace{3em}

\normalsize

${}^a$Yukio Nemoto,
${}^b$Masakiyo Kitazawa and
${}^c$Teiji Kunihiro

\vspace{1.2em}

{\em  ${}^a$Department of Physics, Nagoya University, Nagoya 464-8602, Japan}

{\em  ${}^b$Department of Physics, Osaka University, Toyonaka, Osaka 560-0043, 
Japan}

{\em  ${}^c$Yukawa Institute for Theoretical Physics, Kyoto University,
Kyoto 606-8502, Japan}

\end{center}

\begin{abstract}
We investigate the quark spectrum near but above the critical temperature
of the chiral transition, taking into account the precursory soft modes.
It is found that there appear novel excitation spectra of quasi-quarks
and quasi-antiquarks with a three-peak structure.
By a detailed analysis on the formation of the three-peak structure
using Yukawa models,
it is shown that the new quark spectra originate from the mixing 
between a quark (anti-quark) and an antiquark hole (quark hole) caused 
by a resonant
scattering of the quasi-quark with the soft modes which have a small
but finite excitation energy with a small width near the critical 
temperature.

\end{abstract}


\baselineskip 1.3em

\section{Introduction}

Recently, it is believed that the quark-gluon plasma (QGP) just above 
the chiral and deconfinement phase transitions is an unexpectedly
strongly interacting system, which is mainly based on 
the facts that the created matter at RHIC behaves like a perfect 
fluid \cite{Arsene:2004fa}
and that some mesonic bound states of heavy quarks can survive
above the critical temperature ($T_c$) from Lattice QCD studies
\cite{Asakawa:2003re}.
Because the fundamental degrees of freedom in QGP are quarks and gluons,
it is also important to study their properties in such a strongly interacting
system.
Here, we investigate the quark spectrum just above $T_c$ of 
the chiral transition, focusing on the precursory soft modes.
It is known that the soft modes exist over a wide range of
temperature above $T_c$ owing to a strong coupling nature between 
quarks \cite{Hatsuda:1985eb}.
In this work, we show that they affect the quark spectrum significantly 
in a region just above $T_c$.
In particular, it is shown that 
the quark spectrum form three-peaks at low energy and
low momentum region near $T_c$ \cite{Kitazawa:2005mp}.

In Sec.\ref{NJL}, using a chiral effective model,
we show that the quark spectrum near $T_c$ 
forms the three-peak structure owing to a coupling with the
soft modes.
Such a three-peak structure is seen more clearly if the soft modes
are replaced by an elementary massive boson \cite{Kitazawa:2006zi}.
In Sec.\ref{YKW}, a detailed analysis on the quark spectrum is given
using Yukawa models with a massless quark and a massive scalar
(pseudoscalar) or vector (axialvector) boson.
We elucidate the mechanism of the formation of
the three-peak structure through which the new quark collective
excitations are realized in terms of the Landau damping of a quark
(an antiquark) induced by scattering with the thermally excited
boson, which gives rise to mixing and hence a level repulsion
between a quark (antiquark) and an antiquark hole (quark hole).
Brief summary and concluding remarks are given in Sec.\ref{sum}.

\section{Quark spectrum near chiral transition} \label{NJL}

To describe the chiral transition and the fluctuations,
we first employ the two-flavor Nambu--Jona-Lasinio model
in the chiral limit
\begin{eqnarray}
  \mathcal{L}=\bar{\psi} i \partial \hspace{-0.53em} / \psi
  + G_S [(\bar{\psi} \psi)^2 + (\bar{\psi}i\gamma_5\vec{\tau}\psi)^2],
\end{eqnarray}
with the coupling constant $G_S=5.5$ GeV${}^{-2}$ 
and the three dimensional cutoff $\Lambda=631$ MeV taken from Ref. 
\cite{Hatsuda:1985eb}. 
This model gives a second order phase transition at $T_c=193.5$MeV
for vanishing quark chemical potential.

\begin{figure}[t]
\centering
\includegraphics[width=0.45\textwidth]{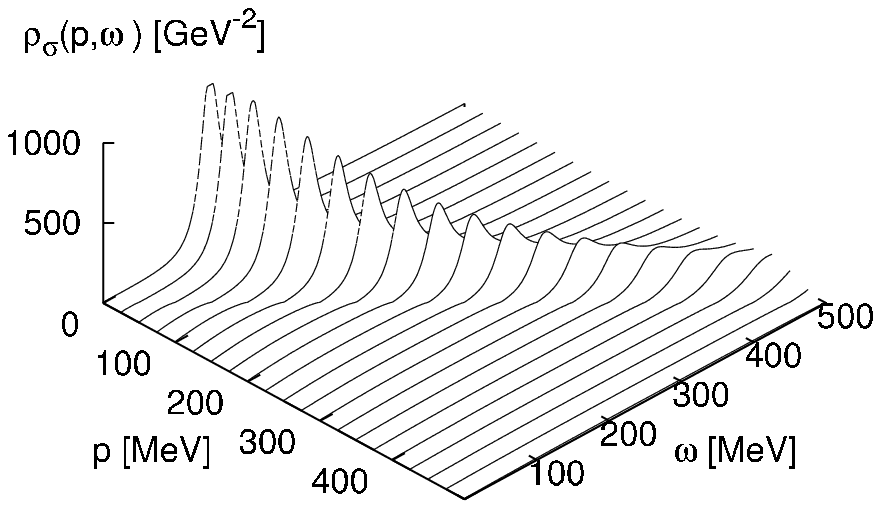}
\includegraphics[width=0.45\textwidth]{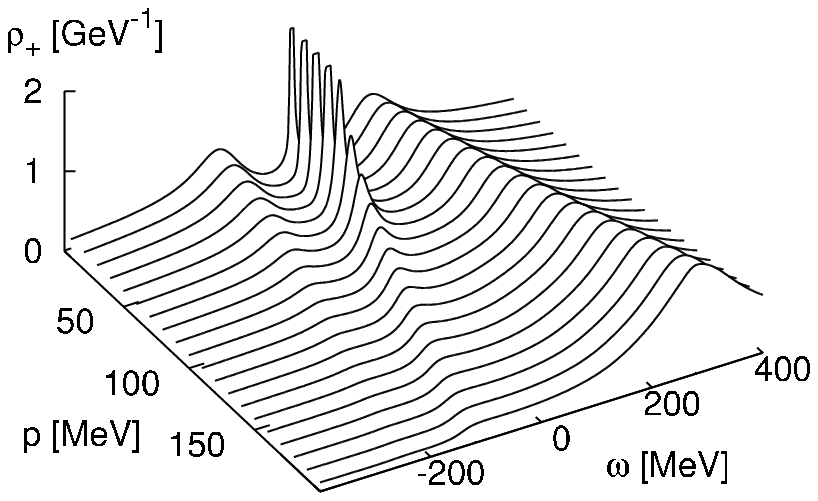}
\caption{The spectral function $\rho_\sigma\equiv\rho$ in the quark-antiquark
channel (left) and 
the quasi-quark spectral function $\rho_+$ (right)
for $\mu=0$ and $\varepsilon\equiv(T-T_c)/T_c=0.1$.}
\label{fig:imd}
\end{figure}

The fluctuations of the chiral condensate are described by
the quark-antiquark Green function in the
random phase approximation,
\begin{equation}
  \mathcal{D}(\bm{p},\nu_n)=-\frac{1}{1/(2G_S)+
  {\cal Q}(\bm{p},\nu_n)},
\end{equation}
where 
$\nu_n=2\pi nT$ is the Matsubara frequency for bosons and
${\cal Q}(\bm{p},\nu_n)$ is the undressed 
quark-antiquark polarization function at one-loop.
To evaluate strengths of the fluctuations, we employ the spectral function,
$\rho$, given by
\begin{equation}
  \rho(\bm{p},\omega) = 
  -\frac{1}{\pi}{\rm Im}{\cal D}(\bm{p},\nu_n)|_{i\nu_n=i\omega+i\eta}.
\end{equation}
We show 
$\rho(\bm{p},\omega)$ near $T_c$ in the left panel of Fig. \ref{fig:imd}.
One can see that there appear pronounced peaks which denote
the precursory soft modes \cite{Hatsuda:1985eb}.
The peak positions of the modes are approximately expressed
as
$\omega\simeq \pm \sqrt{m_\sigma^{*}(T)^2+\bm{p}^2}$.
A $T$-dependent `mass' $m_\sigma^*(T)$ becomes smaller as $T$ approaches $T_c$,
which means the softening at $T_c$.

The effect of the soft modes on the quark spectrum is incorporated in 
the quark self-energy in a non-selfconsistent way,
\begin{equation}
  \tilde{\Sigma}(\bm{p},\omega_n) =
  -4T\sum_m \int \frac{d^3q}{(2\pi)^3} {\cal D}(\bm{p}-\bm{q},
  \omega_n-\omega_m){\cal G}_0(\bm{q},\omega_m),
\end{equation}
where ${\cal G}_0(\bm{q},\omega_m)$ is the free quark propagator with
$\omega_n=(2n+1)\pi T$ being the Matsubara frequency for fermions.
The quasi-quark and quasi-antiquark spectral functions,
$\rho_\pm(\bm{p},\omega)$, are obtained from the retarded self-energies,
$\Sigma^R_\pm(\bm{p},\omega)=(1/2){\rm Tr}[\Sigma^R\gamma^0 \Lambda_\pm]$,
respectively, i.e. 
\begin{equation}
\rho_\pm(\bm{p},\omega)=-\frac{1}{\pi}{\rm Im}[\omega\mp|\bm{p}|-
\Sigma^R_\pm(\bm{p},\omega)]^{-1},
\end{equation}
 with the analytic continuation 
$\Sigma^R(\bm{p},\omega)=
\tilde{\Sigma}(\bm{p},\omega_n)|_{i\omega_n=\omega+i\eta}$ and the projection
operators
$\Lambda_\pm=(1\pm \gamma^0 \bm{\gamma}\cdot\bm{p}/|\bm{p}|)/2$.

The quasi-quark spectral function $\rho_+$ for $\mu=0$ MeV and
$\varepsilon\equiv(T-T_c)/T_c=0.1$ is plotted in the right panel of Fig. \ref{fig:imd}.
We see a clear three-peak structure at low momentum, which
exists even at $\varepsilon=0.2$ \cite{Kitazawa:2005mp}.
Although not shown in the figure, the quasi-antiquark spectrum, $\rho_-$,
has also a three-peak structure for a relation,
$\rho_-(\bm{p},\omega)=\rho_+(\bm{p},-\omega)$.
The mechanism of the appearance of the three-peak structure in
$\rho_+$ is as
follows:
The imaginary part of $\Sigma^R_+(\bm{0},\omega)$ has two peaks
at nonzero values of $\omega$, which means that there exist two
large damping modes of the quasi-quark there.
From a kinematical consideration which is explained in detail
in the following section, 
we see that one is a collision
of a thermally excited antiquark and the quasi-quark creating the soft
mode, and the other is a collision of the quasi-quark and the soft mode
creating an on-shell quark.
Both the processes are interpreted as a Landau damping of the quasi-quark.
The point is that the quasi-quark is a mixed state between quarks and
`antiquark-holes' which are annihilation of thermally excited antiquarks
and have the positive quark number.
Then, these damping modes cause a mixing between quarks and antiquark-holes.
This mixing mechanism can be described in terms of the
resonant scattering as in the case of the color superconductivity
\cite{Kitazawa:2005pp},
although a crucial difference arises owing to the different nature of the
soft modes.
We can show that a coupling with the soft mode with 
{\it a nonzero mass $m_\sigma^*(T)$} is
essential for the appearance of the three-peak structure in the quark spectrum,
as will be explained below in Yukawa models.

\section{Quark spectrum in Yukawa models} \label{YKW}

As mentioned above, the soft modes of the chiral transition have the character
of a well-defined elementary boson with a mass $m_\sigma^*(T)$ and a small
width near $T_c$.
Thus it is seen that a coupling with an elementary massive boson is 
favorable for the three-peak structure of the quark spectrum.
In this section, we show that a system composed of a massless quark and a 
massive boson, as described by Yukawa models, exhibits the three-peak
structure in the quark spectral function \cite{Kitazawa:2006zi}.

\subsection{Scalar (pseudoscalar) boson}

We first investigate the spectrum of a massless quark coupled with a massive 
scalar (pseudoscalar) boson at finite $T$ in a Yukawa model:
\begin{equation}
    \mathscr{L} = \bar{\psi} (i \partial\sla - g \phi) \psi
  + \frac12 \left( \partial_\mu \phi \partial^\mu \phi - m^2 \phi^2 \right).
\label{eq:lag}
\end{equation}
$g$ is the coupling constant and $m$ is the boson mass.

The quark self-energy in the imaginary time formalism
at the one-loop order is expressed as
\begin{eqnarray}
\tilde\Sigma ( \bm{p},i\omega_m )
= -g^2 T \sum_n \int \frac{ d^3\bm{k} }{ (2\pi)^3 }
{\cal G}_0 ( \bm{k},i\omega_n )
{\cal D}( \bm{p}-\bm{k} ,i\omega_m-i\omega_n )
\label{eq:tildeSigma}
\end{eqnarray}
where 
${\cal G}_0 ( \bm{k},i\omega_n ) 
= [ i\omega_n \gamma^0 - \bm{k}\cdot\bm{\gamma} ]^{-1}$
and 
${\cal D} ( \bm{k},i\nu_n ) = [ (i\nu_n)^2 - \bm{k}^2 - m^2 ]^{-1}$
are the Matsubara Green functions for the free quark and scalar boson,
respectively, and
$i\omega_n = (2n+1)\pi T$ and $i\nu_n = 2n\pi T$ are the
Matsubara frequencies for the fermion and boson, respectively.

For the coupling with a pseudoscalar boson, a factor
$i\gamma_5$ is added in both sides of ${\cal G}_0$ in 
eq.~(\ref{eq:tildeSigma}).
Such a factor is, however, canceled out because it anti-commutes with
${\cal G}_0$ and thus the self-energy becomes the same form 
as eq~(\ref{eq:tildeSigma}).
Therefore, the following results and discussion in this section hold 
also for the coupling with a pseudoscalar boson.

\begin{figure}[t]
\begin{center}
\includegraphics[width=0.54\textwidth]{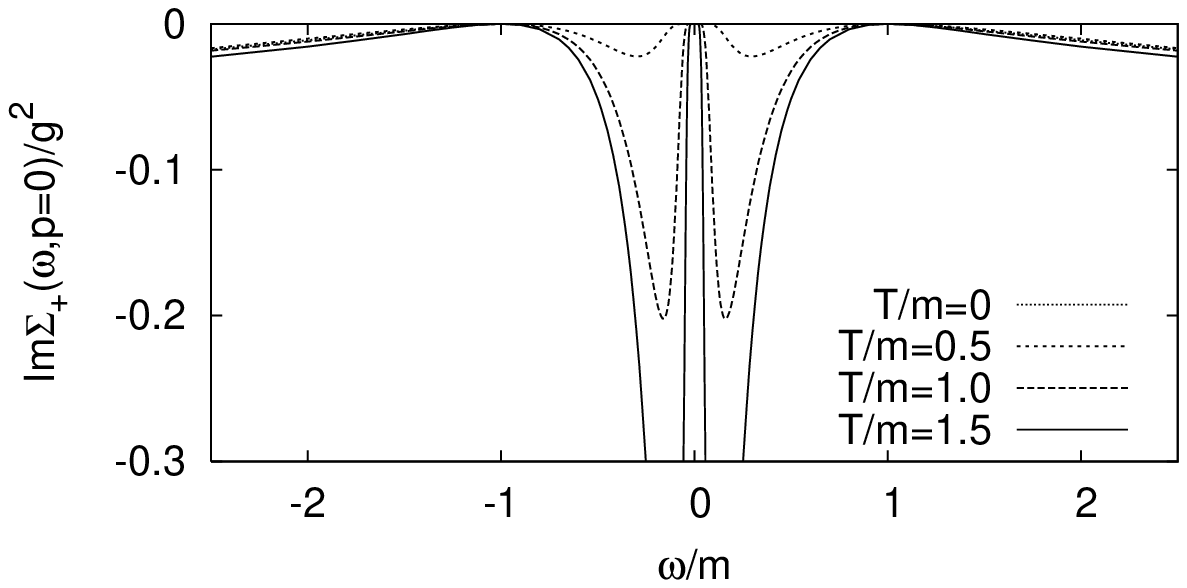}
\includegraphics[width=0.4\textwidth]{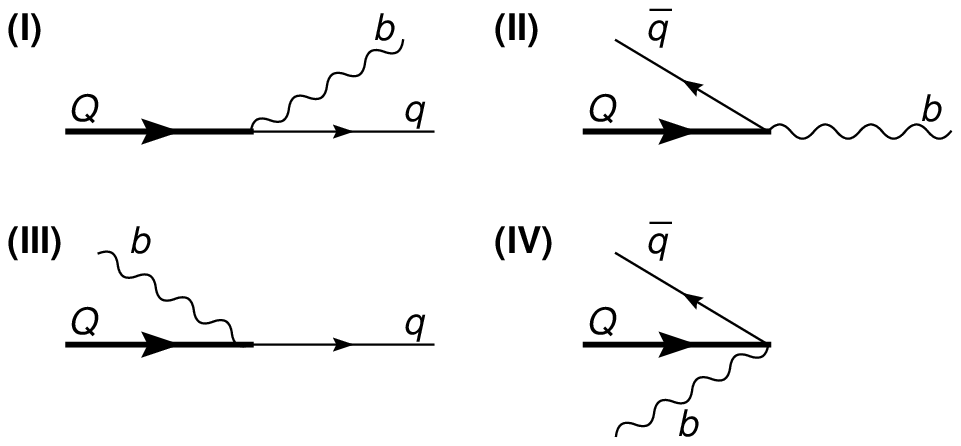}
\caption{
Im$\Sigma_\pm(\bm{p}=0,\omega)$ for several temperatures (left) and
the kinetic processes in Im$\Sigma_\pm(\bm{p},\omega)$.
$Q$ represents the quasi-quark, $q$ on-shell free quarks and 
$b$ the on-shell scalar boson. 
The incident on-shell particles are thermally excited 
particles.
}
\label{fig:ImSigma_p0}
\end{center}
\end{figure}

After the Matsubara summation and the analytic continuation, the
retarded self-energy is obtained.
In the left panel of Fig.~\ref{fig:ImSigma_p0},
we show Im$\Sigma_+ (\bm{p}=0,\omega)/g^2$, which is independent of
$g$.
We see that it vanishes at $\omega=0,\pm m$, irrespective of $T$.
The vanishing decay rate at $\omega=\pm m$ is owing to
the suppression of the phase space;
the energy-momentum conservation requires
 the zero momentum of the on-shell (anti)quark
for each process at $\omega=\pm m$.
Around $\omega=0$, on the other hand, 
the distribution functions suppress the decay rate,
because the on-shell energies of the boson and the quark
 go to infinity as $\omega\to 0$.

At finite $T$, the Landau damping, the processes (II) and (III)
in the right panel of Fig. \ref{fig:ImSigma_p0},
comes to play and Im$\Sigma_+ (\bm{0},\omega)$ have
supports in the regions $-1<\omega/m<0$ for (II) and $0<\omega/m<1$
for (III).
These two peaks in these regions grow  rapidly as $T$ is raised, 
and around $T\simeq m$, the Landau damping is dominant.

\begin{figure}[t]
\begin{center}
\includegraphics[width=0.45\textwidth]{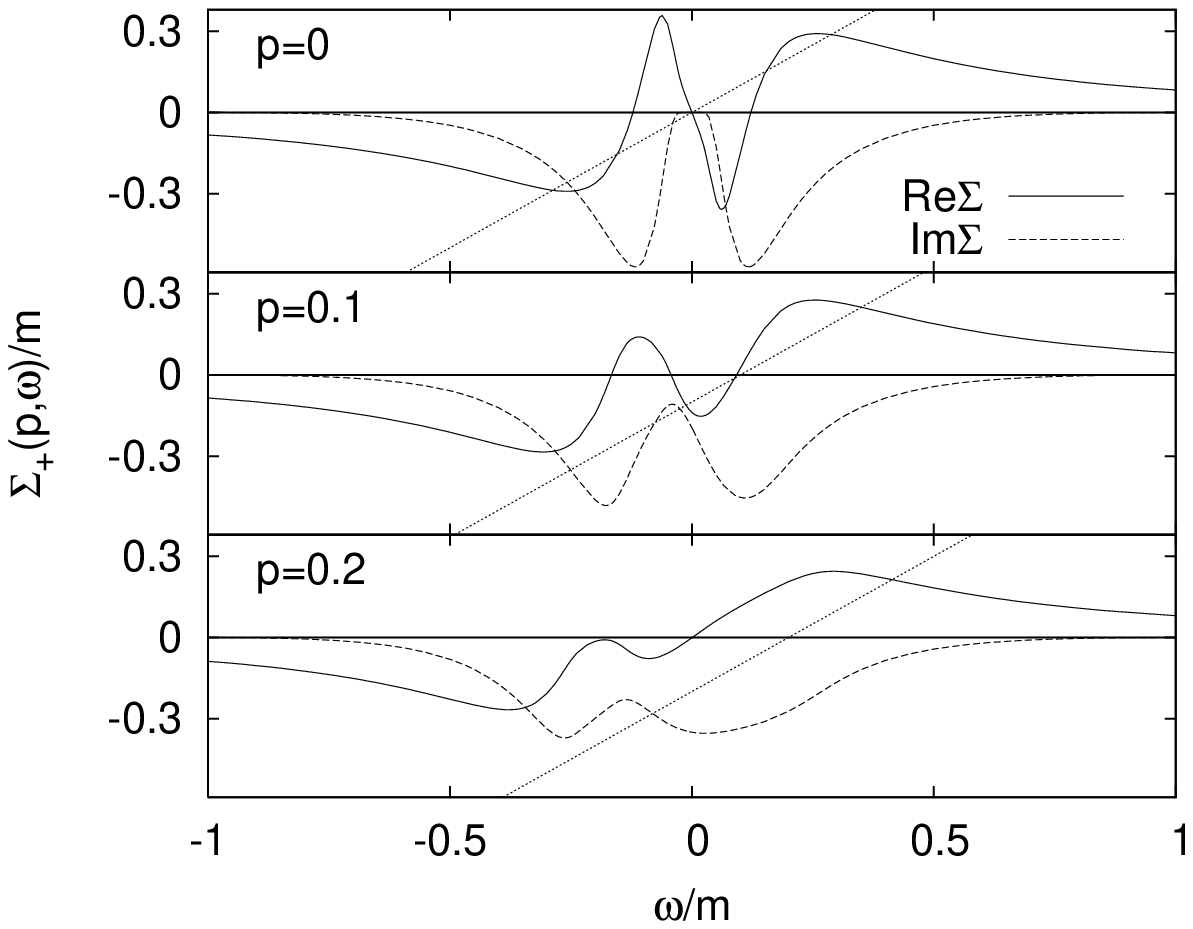}
\includegraphics[width=0.52\textwidth]{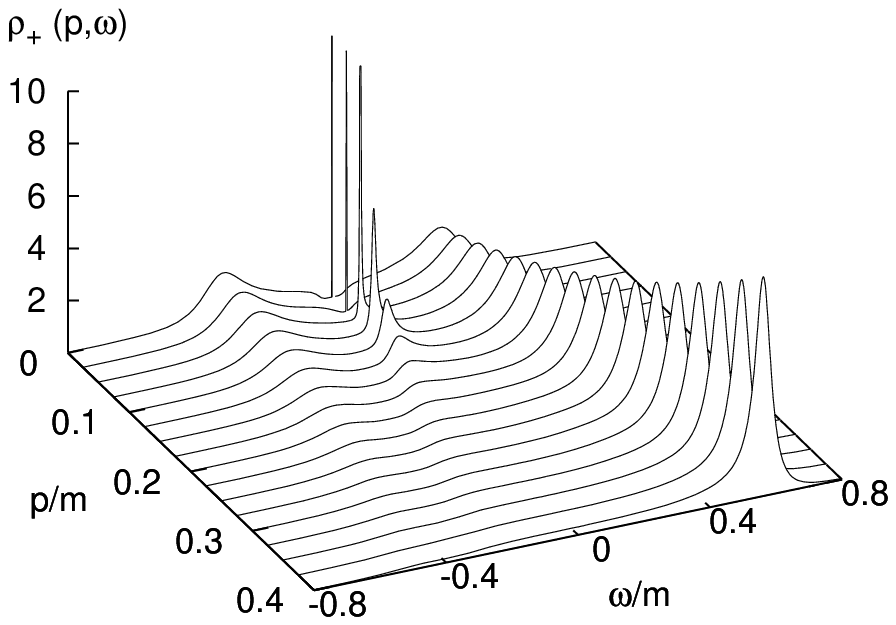}
\caption{
The quark self-energy $\Sigma_+(\bm{p},\omega)$ (left)
and the quark spectral function $\rho_+(\bm{p},\omega)$ (right) 
for $T/m=1.4$.
The dotted straight line in the left panel denotes $\omega-|\bm{p}|$.
}
\label{fig:self}
\end{center}
\end{figure}

The corresponding quark spectral function around for $T/m=1.4$ 
in the right panel of Fig.~\ref{fig:self} shows
a peculiar behavior.
There appear broad peaks 
both in the positive- and negative-energy regions 
as well as the narrow peak around the origin for lower momenta:
Thus the spectral function in the low-momentum region
has \textit{three peaks}.
To understand the behavior of $\rho_+( \bm{p},\omega )$,
we show $\Sigma_+( \bm{p},\omega )$ 
for $T/m=1.4$ 
in the left panel of Fig.~\ref{fig:self}.
We see that 
the oscillatory change of the real part around 
$\omega\sim 0$ is seen, reflecting the peaks in the imaginary part.
It is noted that in general,
a sharp peak in the spectral function can be formed
at $(\omega,\bm{p})$ if
the quasi-dispersion relation, $\omega-|\bm{p}|-{\rm Re}\Sigma_+=0$,
and a relation ${\rm Im}\Sigma_+/{\rm Re}\Sigma_+\ll1$
are hold.
In order to find out the quasi-dispersion relation from Fig.~\ref{fig:self},
we draw the line $\omega-|\bm{p}|$.
For vanishing momentum, we see that there appear five crossing points of
Re$\Sigma_+$ and $\omega-|\bm{p}|$.
The crossing points with the second and fourth largest $\omega$, however,
are located at the energies of the peak of 
$|{\rm Im}\Sigma_+(\bm{p},\omega)|$ 
and hence the spectral function does not form a peak there.
The other three crossing points correspond to the three peaks in the
spectral function.
For large momenta, the number of the crossing points decreases 
and eventually only one crossing point remains.

At higher temperature, the strength of the peak around the origin
is getting weaker, and eventually there remain the other two peaks which
correspond to the normal quasi-quark and the plasmino in the HTL
approximation.

To understand the physical mechanism responsible for
the three-peak structure of $\rho_+( \bm{p},\omega )$,
we first recall that 
there develop two peaks in ${\rm Im}\Sigma_+$ 
which correspond to the decay processes 
(II) and (III).
The process  (II) is 
the annihilation process of the incoming quark $Q$ and the thermally excited
antiquark into a boson in the thermal bath,\,
$Q+\bar{q}\to b$,\, and  its inverse process.
Two remarks are in order here.
First, the disappearance of an anti-quark 
implies the creation of a `hole' in the 
thermally excited anti-quark distribution\cite{Weldon:1989ys}.
Second, the creation of bosons in a thermal bath
is enhanced in comparison with the
case in vacuum by a  statistical factor of $1+n$, 
which becomes large when $T$ is comparable to $m$. 
Thus, we see that the process (II) causes a virtual mixing 
between the quark and `anti-quark hole' states
and as a result, the mixing is enhanced when $T/m\sim 1$. 

The process (III) is another decay process of a quasi-quark
state $Q$, which is now understood to be
a mixed state of quarks and antiquark-holes,
into an on-shell quark 
via a collision with a thermally excited boson:
$Q+ b \to q$ and its inverse process.
These processes again give rise to a mixing of a quasi-quark and
an anti-quark hole state.

\begin{figure}[t]
\begin{center}
\includegraphics[width=140pt]{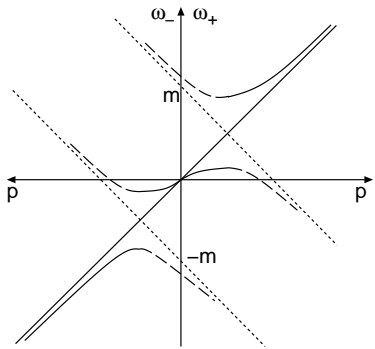}
\caption{Typical peak position of the
spectral functions in the case of level mixing away from the origin.
 The long dashed curves show that the strength of the spectrum
is becoming weaker.
The solid line represents the free quark and antiquark 
dispersion relations, and
the dotted line the antiquark hole and quark hole dispersion
relations.
Level repulsion takes place at the intersection point of these two lines.}
\label{fig:reso}
\end{center}
\end{figure}

The mechanism for the mixing of the quark and hole state 
can also be characterized as a {\em resonant scattering}
\cite{Kitazawa:2005pp,Janko1997},
which was originally introduced to understand the non-Fermi liquid 
behavior of fermions just
above the critical temperature of the superconducting.
In fact, we have seen 
that the process (II)
includes a scattering process of
the  quark  by a massive boson, thus creating
a hole state in the thermally distributed anti-quark states:
$Q\to \bar{q}_h\, +\, b$.
Such a process is called resonant scattering\cite{Kitazawa:2005pp,Janko1997}.
Note that the most probable
process for finite $T$ involves the lowest energy state of the boson,
 i.e., a rest boson with a energy $m$. The energy conservation law
in the most probable case for the above process is
$\omega_{Q}(\bm{p})+\omega_{\bar{q}}(-\bm{p})= m$, or equivalently,
$\omega_{Q}(\bm{p})= m\, -\, \omega_{\bar{q}}(-\bm{p})$.
This equation actually represents the energy-momentum relation for the
particles involved in the state mixing. Thus we see that the
physical energy spectrum is obtained as a result of the level
repulsion between the energies 
$\omega_{q}(\bm{p})=\vert \bm{p}\vert$ and 
$m\, -\, \omega_{\bar{q}}(-\bm{p})=m-\vert \bm{p}\vert$ 
in the perturbative picture.
This situation is schematically depicted in the
upper right part of Fig.~\ref{fig:reso}.

Similarly, the process (III)
includes the process
$q\to Q +b$, and the energy-momentum conservation law for
this process for the most probable case is
$-\omega_{q}(\bm{p})= -m\, + \omega_{Q}(\bm{p})$.
Thus, the physical energy spectrum is obtained as a result of the level
repulsion between the energies 
$-\vert \bm{p}\vert$ and $-m +\vert \bm{p}\vert$.
This situation is also 
depicted in the lower-right part
in Fig.~\ref{fig:reso}.

We thus find that at temperatures satisfying $T/m \sim 1$,
owing to the finite boson mass,
 the level repulsions occur far from the origin and
two gap-like structures in the quark spectrum are formed
at positive and negative energies,
 as shown in Fig.~\ref{fig:reso}.

At
the high temperature limit, $T\gg m$, or $m/T \sim 0$,
the effect of the boson mass can be ignored,
and the resonant scattering occurs only once at the origin
($\omega=|\bm{p}|=0$),
since the energy levels which are to be repelled
cross only there.
Then the quark spectrum has two peaks in the positive and
negative energy region, which is realized in the HTL
approximation.

It was shown in the previous section that 
the quark spectral function possesses a 
three-peak structure near $T_c$ of the chiral transition
 when the chiral soft mode
is incorporated into the quark self-energy.
Recall that
the soft modes behave like a massive elementary 
boson with a mass $m_\sigma^*(T)$ as $T$ approaches $T_c$, i.e.
$\omega\sim \sqrt{\bm{p}^2+m^*_{\sigma}(T)^2}$,
and hence the quark spectra in such a case
are essentially the same as that treated in this section.
We also note that 
as $T$ is lowered toward $T_c$, $m_\sigma^*(T)$ tends to vanish,
and hence the ratio $T/m_\sigma^*(T)$ becomes large.
Thus the quark spectrum
approaches that in the $T/m\to\infty$ limit of the Yukawa model.

\subsection{Vector (axialvector) boson}

In this subsection, 
we investigate the quark spectrum using a Yukawa model
with a massive vector (axial-vector)
 boson  and show that the  three-peak structure
of the quark spectral function is obtained.

We start from the following Lagrangian, composed of 
a massless quark $\psi$ and vector boson field $V_\mu$:
\begin{eqnarray}
{\cal L}
= \bar\psi ( i\partial\sla - ig \gamma^\mu V_\mu ) \psi 
- \frac14 F_{\mu\nu}F^{\mu\nu} + \frac12 m^2 V_\mu V^\mu,
\label{eq:LagrangianVector}
\end{eqnarray}
with the field strength 
$ F_{\mu\nu} = \partial_\mu V_\nu - \partial_\nu V_\mu $.

At one-loop order, 
the quark self-energy in the imaginary time formalism is 
given by
\begin{eqnarray}
\tilde\Sigma ( \bm{p} , i\omega_m )
= -g^2 T \sum_n \int \frac{ d^3 \bm{k} }{ (2\pi)^3 }
\gamma^\mu {\cal G}_0( \bm{k},i\omega_n ) \gamma^\nu 
{\cal D}_{\mu\nu} ( \bm{p}-\bm{k} , i\omega_m-i\omega_n ),
\label{eq:SigmaVImag}
\end{eqnarray}
with the Matsubara propagator for the massive vector boson,
\begin{eqnarray}
{\cal D}_{\mu\nu}( i\nu_n , \bm{p} )
= - \frac{ g_{\mu\nu} - \tilde p_\mu \tilde p_\nu / m^2 }
{ \tilde p_\mu \tilde p^\mu - m^2 },
\label{eq:Proca}
\end{eqnarray}
where $ \tilde p_\mu = ( i\nu_n , \bm{p} ) $.

 For coupling with an axial-vector boson, as the self-energy
has the same form as Eq.~(\ref{eq:SigmaVImag}),
the following results hold also.

\begin{figure}[t]
\begin{center}
\includegraphics[width=0.5	\textwidth]{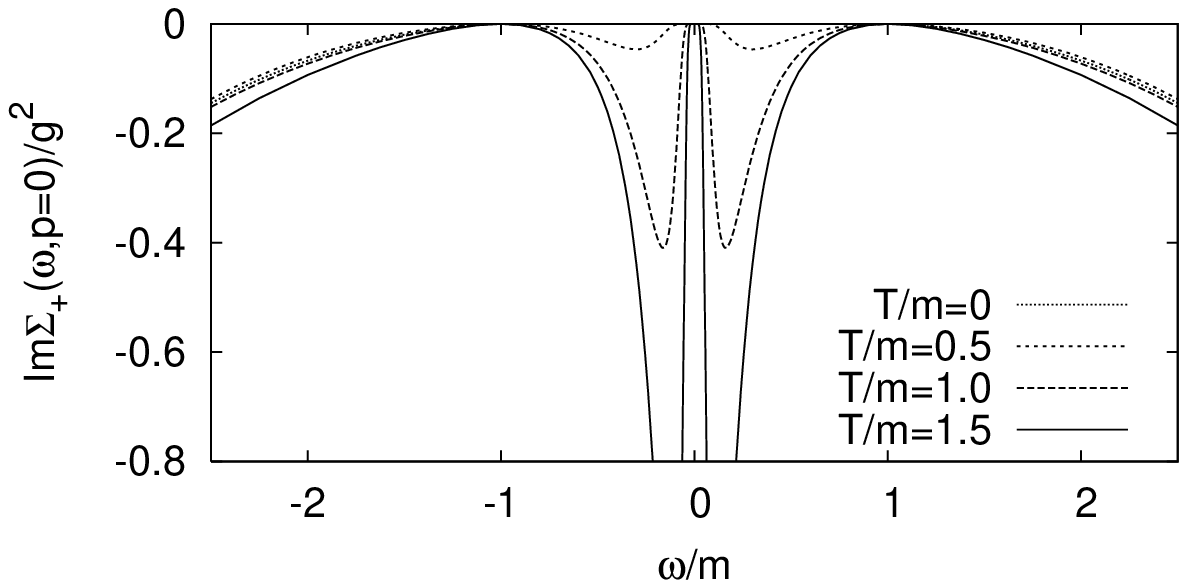}
\includegraphics[width=0.48\textwidth]{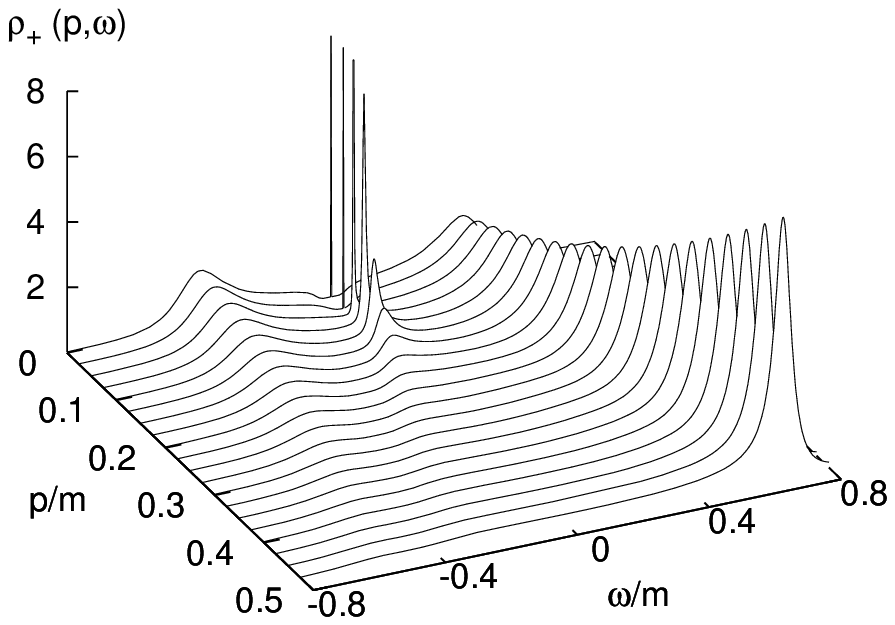} 
\caption{
Im$\Sigma_+ (\bm{p}=0,\omega)$ for several temperatures (left)
and the quark spectral function $\rho_+(\bm{p},\omega)$ for
$T/m=1.2$ (right).
}
\label{fig:ImSigmaP_p0}
\end{center}
\end{figure}

In the left panel of Fig.~\ref{fig:ImSigmaP_p0},
we plot Im$\Sigma_+ (\bm{p}=0,\omega )$ 
at $T/m=0,0.5,1$ and $1.5$.
We fix the coupling constant at $g=1$.
The qualitative features of both parts for $|\omega|/m<1$ are
quite similar to those in the left panel of Fig.~\ref{fig:ImSigma_p0}:
There are two clear peaks in ${\rm Im}\Sigma_+ $ for $|\omega|/m<1$ 
and they grow rapidly as $T$ increases.
For $|\omega|/m>1$, it is seen that 
$|{\rm Im}\Sigma_+(\bm{p}=0,\omega)|$ grows more rapidly 
than in the Yukawa model with the scalar boson shown in 
Fig.~\ref{fig:ImSigma_p0}.

In the right panel of Fig.~\ref{fig:ImSigmaP_p0},
we plot the quark spectral function
for $T/m = 1.2$.
It is seen that there appears a three-peak structure, 
as found in the previous subsection.
This is quite natural, and it would be expected from the
behavior of the self-energy.
A clear three-peak structure is formed at lower values of $T$
than in the scalar boson case, which comes from the difference
of the matrix elements between the scalar and vector boson.

\section{Summary and concluding remarks} \label{sum}

We have investigated the quark spectrum 
near but above the critical temperature of the chiral phase transition
taking the effect of the fluctuations of the chiral
order parameter into account.
We have shown that 
for $\varepsilon\equiv (T-T_c)/T_c \lesssim 0.2$
the quark spectrum has a three-peak structure 
at low frequency and low momentum.
The mechanism of the formation of 
the three-peak structure is based on the fact that  
the soft modes which are composite system
of quark-antiquarks acquire the character 
of the well-defined elementary bosonic excitation
when  $T$ is close to $T_c$.

Then,
we have elucidated
 the essential mechanism of the formation of the three-peak
structure quantitatively 
in Yukawa models composed of a massless quark
and a massive boson with a mass $m$ at finite $T$, in which 
the mass of the boson field is varied by hand.
We have found that
the quark spectral function at low frequency and low momentum
has a three-peak structure at intermediate values of $T$, i.e.
for $T/m\sim 1$, which is independent of the species of the 
massive boson.
Among them, the two peaks have
finite thermal masses and approach the normal quasi-quark and plasmino 
excitations in the HTL approximation in the high $T$ limit, 
while the strength of the other peak around the origin becomes 
weaker and disappears in this limit.

We have discussed the fact that the three-peak structure originates from
the Landau damping, that is, the scattering processes
of a quark and an antiquark hole of a thermally excited antiquark,
and of an antiquark and a quark hole of a thermally excited quark. 
They cause the formation of energy gaps
 in the quark spectrum,
owing to the level mixing between the quark (antiquark) and the hole of 
the thermally excited antiquarks (quarks).
This leads to a level repulsion or gap in the resultant physical spectrum.
These mixings can be understood in terms of the resonant scattering
\cite{Kitazawa:2005pp,Janko1997} of the quasi-quarks off the bosons.
In particular, owing to the mass of the boson, the level repulsion
due to the resonant scattering occurs \textit{twice}
at different points in the energy-momentum plane, 
leading to the three-peak structure of the spectral function.
This contrasts with the case of the quark spectrum coupled to a 
massless boson,
such as a gauge boson, in which the resonant scattering occurs only at
the origin in the energy-momentum plane and then leads to only two peaks,
as in the spectrum in the HTL approximation \cite{Weldon:1989ys}.

The soft modes for the chiral transition behave like a massive 
elementary boson with a mass $m_\sigma^*(T)$ as $T$ approaches $T_c$, i.e.
$\omega	\sim \sqrt{\bm{p}^2+m^*_{\sigma}(T)^2}$,
and hence the quark spectra in such a case
are essentially the same as that treated in the Yukawa models.
We also note that 
as $T$ is lowered toward $T_c$, $m_\sigma^*(T)$ tends to vanish,
and hence the ratio $T/m_\sigma^*(T)$ becomes large.
Thus the quark spectrum
approaches that in the $T/m\to\infty$ limit of the Yukawa models.

In this work,  we have considered only massless quarks.
A finite quark mass should affect the
formation of the three-peak structure in the quark spectral
function. Thus, the
incorporation of the quark mass effect is important for describing the 
chiral transition precisely,
because the order of
the transition, and hence,
the quark spectrum near the critical point are sensitive 
to the quark mass. 
A finite quark mass also leads to a difference
between the scalar (vector) and pseudoscalar (axial-vector) boson 
cases, and may
suppress the three-peak structure.
Detailed study of such a effect is now under way
and will be reported elsewhere\cite{MKKN}.\\

Y.N. is supported by the 21st Century COE Program at Nagoya University
and the JSPS Grand-in-Aid for Scientific Research (\#18740140).
T.K. is supported by a Grand-in-Aid for Scientific Research by Monbu-Kagakusho
of Japan (\#17540240).
This work is supported by the Grand-in-Aid for the 21st Century COE
``Center for Diversity and University in Physics" of Kyoto University.

\end{document}